# High Speed SRT Divider for Intelligent Embedded System

Bhavana Mehta
Dept. of Electronics and Communication,
Nirma University, India
14bec028@nirmauni.ac.in

Jonti Talukdar
Dept. of Electronics and Communication,
Nirma University, India
14bec057@nirmauni.ac.in

Sachin Gajjar
Dept. of Electronics and Communication,
Nirma University, India
14bec028@nirmauni.ac.in

*Abstract*— **Increasing development in embedded systems, VLSI and processor design have given rise to increased demands from the system in terms of power, speed, area, throughput etc. Most of the sophisticated embedded system applications consist of processors; which now need an arithmetic unit with the ability to execute complex division operations with maximum efficiency. Hence the speed of the arithmetic unit is critically dependent on division operation. Most of the dividers use the SRT division algorithm for division. In IoT and other embedded applications, typically radix 2 and radix 4 division algorithms are used. The proposed algorithm lies on parallel execution of various steps so as to reduce time critical path, use fuzzy logic to solve the overlap problem in quotient selection; hence reducing maximum delay and increasing the accuracy. Every logical circuit has a maximum delay on which the timing of the circuit is dependent and the path, causing the maximum delay is known as the critical path. Our approach uses the previous SRT algorithm methods to make a highly parallel pipelined design and use Mamdani model to determine a solution to the overlapping problem to reduce the overall execution time of radix 4 SRT division on 64 bits double precision floating point numbers to 281ns. The design is made using Bluespec System Verilog, synthesized and simulated using Vivado® v.2016.1 and implemented on Xilinx Virtex® UltraScale FPGA board.**

## I. INTRODUCTION

For a long time, the chip industry has depended on the use of personal computers (PC) and its applications for process automation. The heart of the PC consists of various devices that follow Moore's Law. The miniaturization of device geometry has changed the landscape of platform computing. However, personal computer's influence on IC design has slowly been abraded by the flourishment of embedded systems, mobile and portable devices. Embedded systems now dominate every aspect of our life with portable handsets, mobile, tablets, smart IoT based devices etc. [1].

The advent of ubiquitous computing brings with it various challenges: at the technology level, there are several unresolved issues concerning the design and implementation of computing architectures that enable dynamic configuration of ubiquitous services on a large scale. [2]. Portability, power efficiency, computational heterogeneity, throughput, speed are various parameters and areas of concern in the design of embedded systems. A significant amount of change in the architecture and algorithm is required to align with the design constraints and requirements as mentioned earlier. However, even after fifty years of enormous research into altering and changing almost every technology in integrated circuit design, the fundamental arithmetic operations and algebraic structures used in the prevalent embedded systems are still based on the conventional arithmetic units used in personal computers. There is a dire need for increased parallelism and modularity in various operations in an arithmetic unit.

Further, the most complex operation in the arithmetic unit is the division operation, taking up the maximum time and number of cycles. Division is vital and extensively used in almost every computer architecture in a microprocessor. Although its occurrence is rare, the performance of the division operation is the significant contributor to bottlenecks encountered in traditional microprocessor units. Traditionally division in most floating point arithmetic units is implemented mainly by either Goldschmidt's/Newton Raphson algorithm or SRT division method [2].

The Newton Raphson method makes use multiplicative methods utilizing FPU multiplier and requires almost no additional hardware; hence increasing the throughput due to reduction in latency. Whereas the SRT algorithm uses subtractive methods, utilizing a separate hardware for subtraction and shifting in each cycle. Although this increases the hardware and latencies, parallelism is achieved increasing the speed and computational efficiency. With the increase of transistors on a chip as predicted by Moore's law and decrease in price of the same, subtractive methods are widely used as the standalone hardware in most ALUs [3][4]. The derivative of SRT algorithms gives better performance, throughput and computational efficiency. The famous Intel Pentium bug was also due to few incorrect entries in the QST table of SRT radix 4 division algorithm [5].

SRT division is a recursive method producing predicted quotient digit in redundant form at every cycle. The speed depends upon the number of cycles required for the computation to finish. The speed increases as we increase the radix. However, the complexity of quotient digit selection also increases, consequently the size of the quotient selection table increases.

This paper presents a novel approach for the design and implementation of a highly modularized and parallel SRT radix 4 division algorithm where the quotient digit is predicted based on the dividend and then corrected using fuzzy logic. This reduces the size of QST drastically and hence performs better than a traditional SRT radix 4 dividers with or without pre-scaled divisor and dividend. The divider is further interfaced with 64-bit RISC processor, simulated and synthesized on Vivado® and tested on Xilinx Virtex UltraSparc® board.

Section 2 of this paper explains the basic SRT algorithm and its conventional methodology. Section 3 explains the quotient prediction and correction algorithm. Section 4 explains on the fly algorithm and section 5 gives the results obtained followed by a conclusion.

## II. SRT Division Algorithm

The core algorithm of division is a trial and error prone process requiring few initial guesses of a quotient digit followed by subtraction; if the remainder is greater than divisor then the predicted quotient is incorrect and the process is repeated again, discarding the previous result (predicted quotient). In building a computer arithmetic unit, as said earlier, division is the most difficult basic operation to implement in terms of complexity, time and hardware implementation.

SRT division algorithm to implement the division operation in arithmetic unit was given by D. Sweeney, J.E. Robertson [6] and K.D. Tocher [7], in the late fifties simultaneously. Further and the introduction of use of the redundant representations for the remainders was given by D.E. Atkins [4].

SRT division is a recurrent algorithm producing- fixed number of bits, which is equal to the quotient digit in redundant form, at every cycle. For SRT division scheme the recursive relationship is defined by [8]:

$$p_{j+1} = rp_j - q_{j+1}d \quad (1)$$

where the symbols are defined as follows:

$j = $ the recursive index from $0,1,\ldots,m-1$.
$p_j = $ the partial remainder used in the $j^{th}$ cycle
$p_{j+1} = $ the partial remainder used in the $(j+1)^{th}$ cycle
$p_0 = $ the dividend
$p_m = $ the remainder
$q_j = $ quotient digit
$m = $ the number of digits, radix $r$, in the quotient
$d = $ the divisor
$r = $ the radix

For $X$-bit division, where $X = 2n$, it takes $(n-m)$ iterations to compute the final quotient and remainder. The division process needs a quotient selection table to determine the quotient $Q$ and the remaining components. Eq. 1 is partially used to calculate the same. Higher radix SRT division is implemented in such a way that its quotient is selected from a digit set $\{-u \ldots 0, 1 \ldots u\}$.

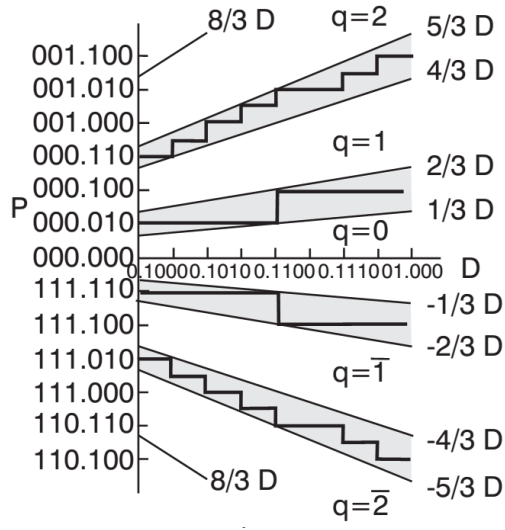

Fig. 1. P-D plot for stair function $(\beta, \alpha) = (4,2)$.

Where u is an integer such that $1/2(r-1) \leq u \leq r-1$, and $r$ is the radix.

The number of bits produced during quotient selection determines the radix i.e $n$ bits quotient digit prediction in each step is radix-$n$ division method. Now, the speed of the operation is directly dependent upon the total number of cycles required for the computation to finish. The speed increases as we increase the radix. However, as we increase the radix - the complexity of quotient digit selection also increases and so does the quotient selection table. In this work the quotient digit is predicted based on the dividend and fuzzy logic; then is corrected if the initial guess was incorrect using non-redundant digits of the dividend. This reduces the size of QST drastically and hence performs better than a traditional SRT radix 4 dividers with or *without* pre-scaled divisor and dividend.

The biggest problem with the traditional approach of SRT division is the prediction of quotient digit in the overlapping regions. These overlapping regions are due to same region corresponding to different coefficient. The quotient digit in the region can be either $q = j$ or $q = j+1$. This implies that we have a choice of values, of both the partial remainder and the divisor that will eventually separate these two adjacent regions. Peter et al. [9] suggested in their paper to use the steps in the overlapping regions and if the ambiguity comes up in the overlapping region the step function will determine the final outcome. This however is extremely complex because the step function determination is not generally for all the overlapping regions. As can be seen from the graph in Fig. 1, the number of steps increases with increasing radix making it more and more complex to execute. Further the selection logic embedded into quotient selection table will become extremely long and implementation will lead to extra hardware, high power consumption and it may violate the critical time violation assumption. This problem can solved using fuzzy logic decision making.

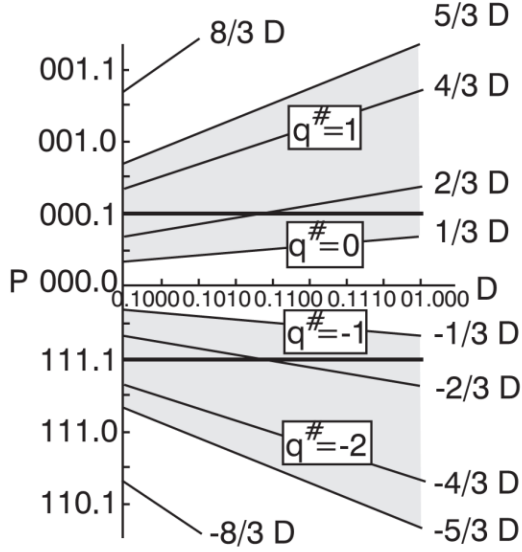

Fig. 2. Uniform overlapped region PD Plot for $(\beta, \alpha) = (4,2)$

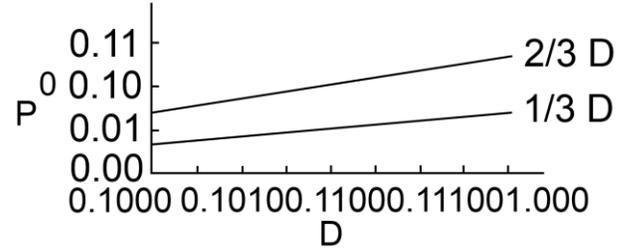

Fig. 3. Common uniform overlapping region for consecutive quotients.

## III. PROPOSED ALGORITHM

The proposed algorithm relies on parallel execution and increased modularity coupled with smarter decision making algorithm (fuzzy logic) in various steps so as to reduce time critical path and hence reduce maximum delay.

Every logical circuit has a maximum delay on which the timing of the circuit is dependent and the path causing the maximum delay is known as the critical path. Our approach uses the previous SRT algorithm methods to make a highly parallel pipelined design to reduce the overall execution time of radix 4 SRT division on 64 bits double precision floating point numbers to $281\ ns$.

The basic concept used in [10], [11] is that the quotient selection algorithm is reduced to two partial algorithms. However, in the presented algorithm a similar methodology is used in a highly modularized and parallel fashion. The overlapping problem is solved by increasing the overlapping area to get a *uniform overlap* within each set of quotients as proposed and illustrated in Fig. 2.

Instead of selecting the correct quotient digit $q$, $-\alpha \leq q \leq \alpha$, which gives rise to a graph as shown in Fig. 1, we estimate a quotient digit $q^{\#}$, $-\alpha \leq q^{\#} \leq \alpha$, such that the actual quotient digit is either $q^{\#}$ or $q^{\#} + 1$, i.e. $q = \{q^{\#}, q^{\#} + 1\}$ [12]. This makes the overlapping region uniform for each consecutive quotients, making the overlapping decision easier. Fig. 3 shows the common uniform overlapping region. For estimated quotient digit $q^{\#}$, and divisor $D$, the upper and lower limits for the corresponding partial remainder are:

$$(-2/3 + q^{\#})D \leq (2/3 + q^{\#} + 1)D \quad (2)$$

Now in the overlapping region the decision problem is solved using fuzzy logic. We studied various fuzzy sets and methods applied in various domains [13-15]. After proper analysis of the input fuzzy set, we figured that there are several ways to determine the output answer based on the inputs, mainly the Mamdani, Larsen, Takagi-Sugeno-Kang, and Tsukamoto inference and aggregation methods are widely used for such problems. We apply Mamdani inference model to the overlapping region.

Let us pick an input value that has membership function in both $q^{\#}$ and $q^{\#} + 1$ region, $P^0 = 0.01$, this will cause both rules to fire. The value 0.01 has a membership of 0.75 in $q^{\#} + 1$ and a membership of 0.25 in $q^{\#}$. Using the Mamdani model and these inputs the resulting aggregate output will be:

$$[I_1(1.25) \wedge 0_1(y)] \vee [I_2(1.25) \wedge 0_2(y)] \quad (3)$$

where input fuzzy set is $I$ and output is $O$.

When all of the possible permutations and combinations have been made, the total output membership function (green), is as shown in Fig. 4. This decision hence will now determine the quotient correction. In a nutshell, the interim quotient $q^{\#}$ and correction quotient is calculated in parallel with new dividends/partial remainder. These are calculated for both the interim quotient and the same incremented one by one as functions outside the main body- partrem1 and partrem2 (partial remainder) respectively.

Further, on the fly algorithm is implemented using a minimalistic approach involving conversion table reduced to four rows and two columns.

### A. Interim Quotient

Here we take interim quotient $Q$ such that $Q \in [-a, a+1]$ which implies the quotient can be either $q$ or $q + 1$. As a consequence range of partial remainder is:

$$P \in [(-2/3 + q), (-2/3 + q + 1)] \quad (4)$$

Since, $q = [-2, 1]$, hence we get the following value of $P$,

$$P \in [(-2/3 - 2), (-2/3 + 1 + 1)]$$

$$P \in [(-8/3), (8/3)]$$

Hence, as discussed in [8], [10-11], the PD plot will change as shown below increasing the overlapping area and hence introducing uniformity. Of the overlapping region as shown in the Fig. 2.

The graph is divided into 4 regions exploiting the uniformity of overlapping regions. This partition is done using three horizontal line partitions as shown in Fig. 2. The first one being X-axis and $c = (000.1)$, $c' = (111.1)$.

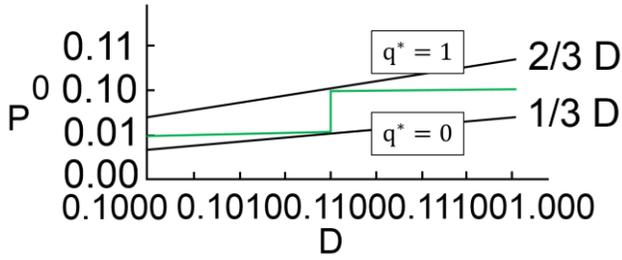

Fig. 4. Total output membership function for consecutive quotients.

$$Q = \begin{cases} 1, & 000.1 \leq P \\ 0, & 0 \leq P \leq 000.1 \\ -1, & 111.1 \leq P \leq 0 \\ -2, & P \leq 111.1 \end{cases} \quad (5)$$

This makes the decision logic of interim quotient low latency and faster than the traditional complex methods. However the decision is incorrect in overlapping regions.

*B. Correction Quotient*

From Fig. 4 it is derived that the correction quotient is a step function, $q^*$ is determined by using comparison constant $C_1=0.01$ for $D$ [0.10, 0.11] and $C_2=(0.10)$ for $D = [0.11, 1.00]$ as shown in Table I.

TABLE I: OVERLAPPING REGION FOR ESTIMATED QUOTIENT

| | Partial Remainder P |
|---|---|
| $P = 0, 1$ | $1/3\ D\ \leq P \leq 5/3\ D$ |
| $P = -1, 0$ | $-2/3\ D\ \leq\ P \leq 2/3\ D$ |
| $P = -2, -1$ | $-5/3\ D\ \leq\ P\ \leq 1/3D$ |

Hence, from the graph and table, the following equation for the correction quotient can be derived using combinational deduction:

$$q^* = S'(P_0 + P_{-1} + P_{-2}d'_{-2}) \quad (6)$$

*C. Partial Remainder*

Since the quotient can be either $Q$ or $Q + 1$ (which can only be determined after the quotient correction is calculated), we in parallel calculate two possible remainders $P_0$ and $P_1$. This is done so that the critical path can be reduced by parallel calculated both the possible remainders and then calling the appropriate one after the correction quotient is calculated:

$$P_0 = P + q^*D \quad (7)$$
$$P_1 = P - (q^* + 1)D \quad (8)$$

The parallelism does not affect the critical path and hence increases the speed.
If the quotient correction was 0 then P0 is used, otherwise P1 is used.

*D. On the Fly Algorithm*

Unlike restoring or non-restoring division methods, SRT division produces redundant digits as a result which needs to be converted into a non-redundant digital set. This can be done by subtracting the positionally weighted negative digit quotient from positive counterpart. However, it will require a carry propagation subtraction unit which can be eliminated if we use on the fly algorithm. In this case the extra hardware will be required [16]. However, in our implementation we have the traditional algorithm which has been shortened to a reduced look up table which runs parallel with each iteration.

As a result, our derived table is superior in terms of time for parallel computation of non-redundant digit is lesser than the time taken for each iteration and hence it the critical path isn't affected. The implementation requires two registers and separating the digit vector to $A[k]$ and $B[k]$ for positive and negative quotient digit respectively [17]. For specifically radix 4 division with each cycle two quotient digits are being predicted. So the quotient digit obtained of base 4 will be converted into non redundant representation resulting in appending two binary digits to $A[k]$ and $B[k]$ i.e., Appending them with $(a1, a0)$ and $(b1, b0)$ respectively. The values for the same is given in Table II.

TABLE II: ON THE FLY ALGORITHM REDUCED TABLE

| Final Quotient | $A[k+1] = (A[k], a1, a0)$ | $B[k+1] = (B[k], b1, b0)$ |
|---|---|---|
| 0 | $(A[k], 0, 0)$ | $(B[k], 1, 1)$ |
| 1 | $(A[k], 0, 1)$ | $(B[k], 0, 0)$ |
| -1 | $(A[k], 1, 1)$ | $(B[k], 0, 0)$ |
| 2 | $(A[k], 1, 0)$ | $(B[k], 0, 1)$ |
| -2 | $(A[k], 1, 0)$ | $(B[k], 1, 1)$ |

IV. RESULTS AND COMPARISON

The algorithm was made for 64 bit floating point double precision numbers. The developed architecture in this paper was designed in Bluespec System Verilog in Linux-Ubuntu-14.04. The design was integrated, simulated and synthesized by Vivado v.2016.1and implemented on Xilinx Virtex UltraScale® FPGA board. The critical time came out to be 210 ns which is significantly lesser than earlier proposed algorithms. Synthesis and simulation showed significant improved results of the architecture in the critical time period, speed, power and area

```
    Site Type          | Used | Fixed | Available | Util%
----------------------+------+-------+-----------+------
CLB LUTs*             | 1879 |     0 |    537600 |  0.35
  LUT as Logic        | 1879 |     0 |    537600 |  0.35
  LUT as Memory       |    0 |     0 |     76800 |  0.00
CLB Registers         |  283 |     0 |   1075200 |  0.03
  Register as Flip Flop | 283 |     0 |   1075200 |  0.03
  Register as Latch   |    0 |     0 |   1075200 |  0.00
CARRY8                |   43 |     0 |     67200 |  0.06
F7 Muxes              |    5 |     0 |    268800 | <0.01
F8 Muxes              |    0 |     0 |    134400 |  0.00
F9 Muxes              |    0 |     0 |     67200 |  0.00
```

Fig. 5. Xilinx VirtexUltraScale® FPGA board resource utilization for super pipelined On the Fly SRT Algorithm.

consumed compared to other algorithms proposed. Further the frequency obtained is approximately 1.5 GHz. The board resource utilization is given in the Fig. 5.

## V. SCOPE AND FUTURE WORK

The algorithm can further be extended to higher radices to implement on higher end processors like S class, M class, H class and T class. For lower radices i.e., Radix 2 which is used in lower end embedded systems, a trade-off will be made and taking into consideration the requirements of the system radix 2 with parallelism and additional hardware will be implemented. By changing the typical SRT algorithm with the variant of parallel SRT in FPUs used in embedded system the speed will increase considerably.

## VI. CONCLUSION

This paper presents a novel approach of introducing parallelism in all the paths with latency lesser than the critical path. This approach reduces the critical time and hence worst slack greatly; hence giving faster and more efficient FPU. Furthermore the proposed algorithm implemented on FPUs can be used in arithmetic units for SoC in smart and sophisticated embedded systems and processors or for Intellectual property in a block of logic of VLSI systems

ACKNOWLEDGEMENT

Authors acknowledge the support provided by RISE (Reconfigurable and Intelligent Systems Engineering) Group, Dept. of Computer Science and Engineering, Indian Institute of Technology, Madras (IIT-M).